\documentclass[runningheads]{llncs}
\usepackage{graphicx}
\usepackage{xcolor}
\usepackage{indentfirst}
\usepackage[subtle,margins=normal,leading=normal]{savetrees}
\usepackage[colorlinks,citecolor=green]{hyperref}
\usepackage{floatrow}
\begin{document}
\title{The Power of Proxy Data and Proxy Networks\\for Hyper-Parameter Optimization\\in Medical Image Segmentation}
\titlerunning{The Power of Proxy Data and Proxy Networks}
% \author{Vishwesh Nath\inst{1}\orcidID{0000-1111-2222-3333}
\author{Vishwesh Nath\inst{1} \and
Dong Yang\inst{1} \and
Ali Hatamizadeh\inst{1} \and
Anas A. Abidin\inst{1} \and
Andriy Myronenko\inst{1} \and
Holger R. Roth\inst{1} \and
Daguang Xu\inst{1}}
%index{Nath, Vishwesh}
%index{Yang, Dong}
%index{Hatamizadeh, Ali}
%index{Abidin, Anas}
%index{Myronenko, Andriy}
%index{Roth, Holger}
%index{Xu, Daguang}

\institute{NVIDIA\inst{1}}

\authorrunning{V. Nath et al.}
%\institute{Princeton University, Princeton NJ 08544, USA \and
%Springer Heidelberg, Tiergartenstr. 17, 69121 Heidelberg, Germany
%\email{vnath@nvidia.com}\\
\maketitle              % typeset the header of the contribution

\begin{abstract}
Deep learning models for medical image segmentation are primarily data-driven. Models trained with more data lead to improved performance and generalizability. However, training is a computationally expensive process because multiple hyper-parameters need to be tested to find the optimal setting for best performance. In this work, we focus on accelerating the estimation of hyper-parameters by proposing two novel methodologies: proxy data and proxy networks. Both can be useful for estimating hyper-parameters more efficiently. We test the proposed techniques on CT and MR imaging modalities using well-known public datasets. In both cases using one dataset for building proxy data and another data source for external evaluation. For CT, the approach is tested on spleen segmentation with two datasets. The first dataset is from the medical segmentation decathlon (MSD), where the proxy data is constructed, the secondary dataset is utilized as an external validation dataset. Similarly, for MR, the approach is evaluated on prostate segmentation where the first dataset is from MSD and the second dataset is PROSTATEx. First, we show higher correlation to using full data for training when testing on the external validation set using smaller proxy data than a random selection of the proxy data. Second, we show that a high correlation exists for proxy networks when compared with the full network on validation Dice score. Third, we show that the proposed approach of utilizing a proxy network can speed up an AutoML framework for hyper-parameter search by 3.3$\times$, and by 4.4$\times$ if proxy data and proxy network are utilized together. 

\end{abstract}

\section{Introduction}

Data-driven methods have become the main approach for medical image segmentation based tasks for most imaging modalities such as computed tomography (CT) and magnetic resonance imaging (MRI) \cite{litjens2017survey,tajbakhsh2020embracing}. At the same time, the drive for growing annotated data has been accelerating at an exponential rate \cite{wiesenfarth2021methods}. However, the performance of deep learning \cite{lecun2015deep} methods is critically dependent upon the hyper-parameters (learning rate, optimizers, augmentation probabilities, etc) that are utilized while training the model. Hyper-parameter optimization (HPO) \cite{bergstra2012random} is actively being researched by state-of-the-art (SOTA) approaches such as AutoML~\cite{real2020automl,hutter2019automated}. However, AutoML is usually an extremely computationally expensive process. Furthermore, the computational cost is exacerbated by medical imaging tasks where the data is often high-dimensional, i.e. 3D volumes.

\begin{figure}[!htb]
\begin{center}
    \includegraphics[width=0.7\linewidth]{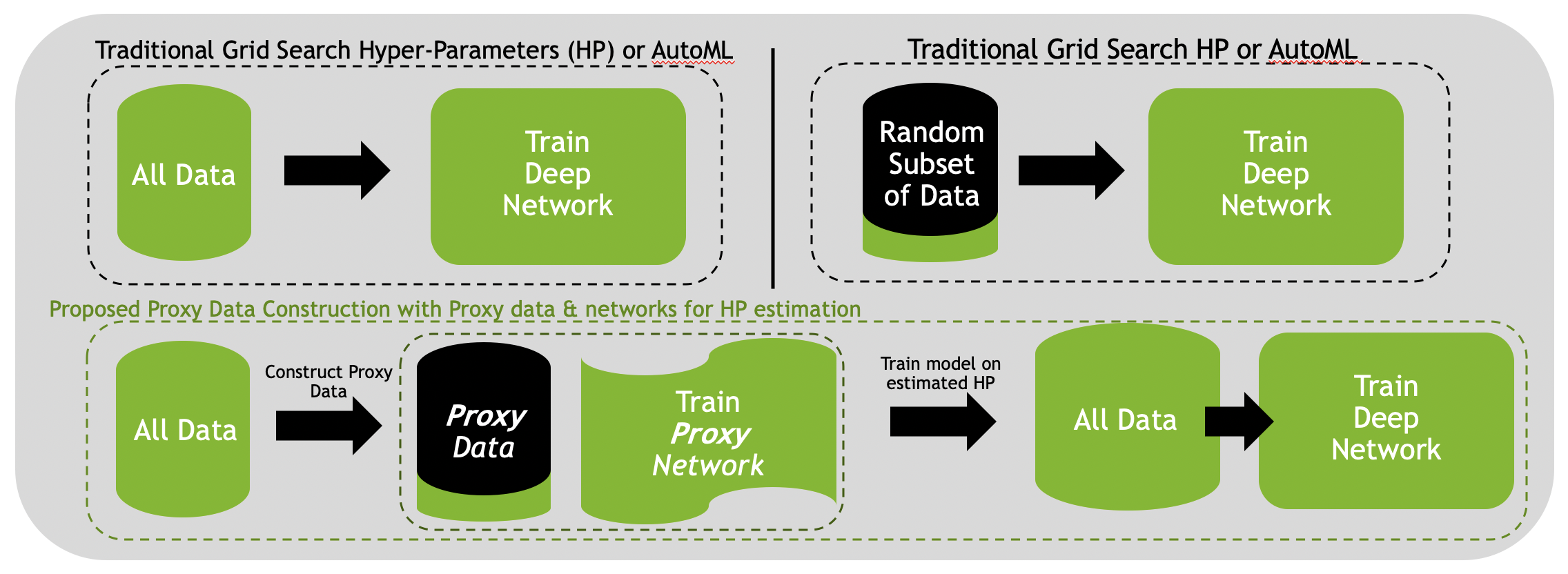}
    \caption{The proposed pipeline of utilizing proxy data and proxy network to accelerate hyper-parameter estimation. Top row: Traditionally, hyper-parameters are either estimated on the full dataset or on a small subset of randomly selected data. Bottom row: Combining proxy networks with proxy data directly lead to speed ups in estimating hyper-parameters.}
    \label{fig1:proxy_idea}
\end{center}
\end{figure}

Our work focuses on reducing the computational expense of HPO and AutoML approaches. In this paper, we propose the construction of a proxy dataset, which is representative of the full dataset yet relatively much smaller in size (see Fig. \ref{fig1:proxy_idea}). Furthermore, we also propose the usage of proxy models, which are essentially smaller networks, yet representative of the larger/complete network structure. Utilizing the proposed proxy data and proxy networks, we show that computational burden of AutoML can be drastically reduced \textit{from a few days to a few hours} and yet be able to estimate hyper-parameters that can lead to SOTA performance. For a robust validation of the approach of proxy data and proxy networks we utilize external validation data (completely withheld datasets) for testing of the trained models from the primary datasets. The approach is tested on CT and MR imaging modalities.

\section{Related Work}

\textbf{Proxy Data \& Networks:} The term \textit{proxy} implies computationally reduced settings \cite{coleman2019selection}. Prior work has shown that proxy datasets can be useful in AutoML, specifically for sub-domains of Neural Architecture Search (NAS) and HPO \cite{park2019data,shleifer2019using}. However, they have only been tested for classification tasks by reduction of class labels or mining of easy and hard samples \cite{shleifer2019using}. Similarly, proxy networks have been associated only with classification tasks \cite{coleman2019selection}. To the best of our knowledge, there is no previous work which has proposed proxy data and proxy networks for medical image segmentation.

\textbf{AutoML:} Since the inception of AutoML \cite{hutter2019automated}, it has become the best way to estimate hyper-parameters and is effectively better than grid search. The prior literature indicates the popularity of AutoML \cite{hutter2019automated,real2020automl} and it has also been effectively introduced for medical image segmentation \cite{yang2019searching,yu2020c2fnas}. While AutoML methods exist, their practical feasibility is of critical concern due to their heavy computational expense, which is a primary motivation for our work.

The proxy data and proxy networks serve as useful and much needed indicators of performance that can drastically speed up the estimation of hyper-parameters while reducing the computational expense.

\subsection{Contributions:}
\begin{enumerate}
    \item We propose the construction of proxy datasets via classical metrics of mutual information and normalized cross correlation. Our work also shows that the choice of metric needs to be paired with the spatial context of the task.
    \item We show that proxy networks can be constructed systematically by reduction of residual blocks, channels and reducing number of levels in the U-net.
    \item We show that our methods can be generalized across different imaging modalities, via evaluation on external validation datasets from which no data were utilized for training.
\end{enumerate}

\section{Method}

\textbf{Proxy Data Selection Strategy:} Consider a dataset $D$ containing a set of datapoints $\{x_1, x_2, .. x_n\}$, where $x_i$ is a single data sample. To estimate the importance of a single datapoint $x_i$, we estimate its utility in relation to other datapoints $x_j$, resulting in a set of paired measures. An example of pairs for $x_1$ would be $\{(x_1, x_1), (x_1, x_2), (x_1, x_3) .. (x_1, x_n)\}$. The mean of the measure is utilized as an indicator of the importance of the datapoint. There are multiple methods that can provide pair-wise measurements of the data. We explore mutual information (MI) (Eq. \ref{eq:mutual_info}) on flattened vectors of the 3D images \cite{nath2020diminishing} and normalized local cross-correlation (NCC) (Eq. \ref{eq:ncc}) in local window size of (9, 9, 9) \cite{zhu2020neurreg} for each pair of data $(x_i, x_j)$ as different variants. 

\begin{equation}
\small
    \mathcal{MI}(x_i, x_j) = \sum_{x_i}\sum_{x_j}P(x_i, x_j)log\frac{P(x_i, x_j)}{P(x_i)P(x_j)}.
    \label{eq:mutual_info}
\end{equation}

Here $P(x_i)$ \& $P(x_j)$ are the marginal probability distributions while $P(x_i, x_j)$ is the joint probability distribution.

\begin{equation}
\small
    \mathcal{NCC}(x_i, x_j) = \frac{1}{\Omega}\sum_{p \epsilon \Omega}\frac{(\sum_{p_i}(x_i(p_i) - \overline{x_i(p)})(x_j(p_i) - \overline{x_j(p)}))^2}{\sum_{p_i}(x_i(p_i) - \overline{x_i(p)})^2 \sum_{p_i}(x_j(p_i) - \overline{x_j(p)})^2}
    \label{eq:ncc}
\end{equation}

Here, $p_i$ is the 3D voxel position within a window around p and $\overline{x_i(p)}$ and $\overline{x_j(p)}$ are local means within the window surrounding the voxel position $p_i$ in $x_i$, $x_j$ correspondingly. $\Omega$ is the voxel coordinate space.

\textit{Task-specific Region of interest}: The acquisition parameters (number of slices, resolution etc.) for different 3D volume scans vary. Therefore, when considering a pair $(x_i, x_j)$, even if the $x_i$ is re-sampled to $x_j$ image size, there is misalignment for the region of interest (ROI) (the organ to be annotated by the model). Hence, we utilize only the task-specific ROI by utilizing the information from the existing label. The selected volume is cropped using the ROI and re-sampled to a cubic patch size.

The data points are ranked by their importance and the ones containing the lowest mutual information or lowest correlation are selected within a given budget $B$. 

\textbf{Proxy Network:} U-net has become the go-to model for medical image segmentation tasks especially when deep learning methods are being utilized \cite{ronneberger2015u}\cite{isensee2019automated}. There are many variants of U-net that were proposed, we use a 5 level U-net with 2 residual blocks per level and skip connections between encoder and decoder blocks (this is our full model). The coarse hyper-parameters for a U-net are therefore: number of channels in first encoder block (successive encoder blocks are multiples of 2), number of residual blocks and the number of levels. To create a proxy network, we first reduce the number of channels to 4 and decrease the residual blocks from 2 to 1. The variants of proxy network are created by decreasing the number of levels to 5, 4 \& 3 (can also be thought as reducing the number of encoding and decoding blocks).

\textbf{AutoML:} We use a recurrent neural network (RNN) that is part of the over-arching reinforcement learning (RL) framework to estimate hyper-parameters \cite{yang2019searching}.

\section{Experiments \& Data}

\subsection{Datasets}

\textbf{CT}: The Spleen task dataset from MSD \cite{simpson2019large} with the segmentation annotation were used. Data was re-sampled to a resolution of $1.5 \times 1.5 \times 2.0$ $mm^3$ and the intensities were normalized in a HU window of [-57, 164]. Random patches of $96 \times 96 \times 96$ with or without the labels were used for training. For inference during validation and testing a patch size of $160 \times 160 \times 160$ was utilized. All 41 volumes were utilized for training and validation for ground truth. 

External validation based testing was done on all the 30 volumes from Beyond the Cranial Vault (BTCV) challenge from MICCAI 2015 \cite{landman2015miccai}. The pre-processing, patch-size for inference were kept consistent as with the first dataset. Please note that all other labels were ignored from BTCV as it is a multi-organ annotated dataset.

\textbf{MR}: The Prostate task from MSD \cite{simpson2019large} with the segmentation annotation were used. The original task includes the labels of transition zone and peripheral zone separately, for this work these two were combined to a single class, for consistency with the secondary dataset. Data was pre-processed using a resolution of $1.0 \times 1.0 \times 1.0$ $mm^3$, intensities were normalized, spatially padded to ensure consistency, random patches of $128 \times 128 \times 48$ were selected with or without the label using a ratio. For inference a patch size of $160 \times 160 \times 160$ were used. All 32 volumes were utilized for training and validation for ground truth.

External validation based testing was done on all 98 volumes from the PROSTATEx dataset \cite{litjens2014computer}. The pre-processing, patch-size for inference were kept consistent as with the first dataset. 

For all pre-processing and training of deep learning models, the MONAI library\footnote{\url{https://monai.io}} was used.

\subsection{Experimental Design}

\textbf{Hyper-Parameter Space:} We explore a hyper-parameter space of four different optimizers $\phi=$\{Adam, RMSProp, Adamax, Novograd\}  at varying learning rates in the set $\delta=$\{0.001, 0.0006, 0.0004, 0.0001\}  for validation of the proxy data selection strategy.

\textbf{Ground Truth:} The ground truth performance of different hyper-parameters was estimated on 7 random splits of training and validation for both MR and CT MSD datasets. For MSD Spleen dataset, training and validations splits are of 32 and 9 sample size. Similarly, for MSD Prostate, sample sizes were 26 and 6. Please note that no data samples from the secondary datasets were used for training. They were only used for testing as external validation datasets.

We aim to answer the following questions with our experimental design.

\noindent\textbf{\textit{Which proxy dataset construction method is the best and how many data points are sufficient for a good proxy?}} The metrics of MI and NCC with or without combinations of task-specific ROI selection are tested across 7 repeats (different initialization of the network, different data splits) where the selected $B$ datapoints are randomly split into training set (50\%) \& validation set (50\%). We also test the baseline of random selection of data. All data selection techniques are tested with $B$=\{4, 6, 8 and 10\} data points in total. To evaluate the best method, we study the correlation (Pearson correlation coefficient) of the Dice scores estimated on the secondary validation dataset by training on proxy data versus the Dice scores of the same validation set but trained with all training data (the ``full'' model). Meanwhile, through an ablative study, we show how much proxy data is needed for it to be a good proxy representing the entire dataset in terms of ground truth Dice scores.

\noindent\textbf{\textit{How shallow can a proxy network be?}}
To evaluate, how small of a proxy network can be used, we systematically decrease the number of levels of the U-Net from 5 to 3 at steps of 1. The Pearson correlation is studied across seven different splits of data which are trained on seven different initializations of the network for proxy networks versus the full model.

\noindent\textbf{\textit{Will proxy data and proxy network lead to relatively closer hyper-parameters when estimated on all data and a full model?}} We utilize the RL AutoML technique which is used to estimate hyper-parameters. Learning rate and the probability for the augmentation of a random shift in intensity are estimated via AutoML. The relative distance between the estimated parameters from proxy data and proxy networks is compared with hyper-parameters searched with a full model and all data. We also show comparisons when the hyper-parameters are estimated with a random selection of data. Once the hyper-parameters are estimated using a proxy technique, a full model with all data is trained for evaluation purposes.

\begin{table*}[!h]
  \centering
  \scriptsize
  \begin{tabular}{p{2.5cm} p{2.0cm} p{2.0cm} p{2.0cm} p{2.0cm}  }
     \hline
     \multicolumn{5}{c}{\textbf{CT | Spleen \ \ \ \ \ \ \ \ \ \ \ \  MRI | Prostate}} \\
     \hline
     Selection Method & MSD Spln Dice & BTCV Dice & MSD Prst Dice & PRSTx Dice  \\\hline
     NCC & $0.8804 \pm 0.0314$\ & $0.8012 \pm 0.0247$ & $0.6180 \pm 0.1044$\ & $0.4059 \pm 0.1326$ \\
     NCC+Labelcrop & $0.9223 \pm 0.0194$\ & $0.6579 \pm 0.0681$ & $0.6647 \pm 0.0955$\ & $0.4165 \pm 0.1190$\\
     MI & $0.8814 \pm 0.0405$\ & $0.7446 \pm 0.0469$ & $0.5365 \pm 0.0792$\ & $0.3401 \pm 0.0706$ \\
     \textbf{MI+Labelcrop} & $0.8580 \pm 0.0440$\ & $\textbf{0.8173} \pm 0.0184$ & $0.6312 \pm 0.0383$\ & $\textbf{0.5567} \pm 0.0754$ \\
     \hline
     Random baseline & $0.8282 \pm 0.0559$\ & $0.7678 \pm 0.0391$ & $0.5619 \pm 0.1318$\ & $0.4298 \pm 0.1406$ \\
     \hline
     Full Dataset & $0.9450 \pm 0.0106$\ & $0.8844 \pm 0.0037$\ & $0.8534 \pm 0.0121$ & $0.7423 \pm 0.0497$ \\
     \hline
  \end{tabular}
  \caption{Summarized mean Dice score for internal validation of MSD spleen, external validation of BTCV, internal validation of MSD prostate, external validation of PROSTATEx across 7 random splits of selected data. NCC represents normalized cross-correlation and MI represents Mutual information. Spleen Dice scores are reported with 23\% usage of full dataset for training. Prostate Dice scores are reported with 31\% usage of full dataset}
  \label{tab:method_select}
\end{table*}

\section{Results}

\textbf{Proxy Data:} It should be noted that the test Dice results should be compared relatively as they have been trained on reduced datasets and the performance is not expected to reflect the highest Dice scores. The best test Dice score was selected from the hyper-parameter search space for the ground truth and the corresponding settings were used for all the data selection methods (listed in Tab.~\ref{tab:method_select}). The method \textit{MI+Labelcrop} (Labelcrop is referred to as the selected ROI based on the label) shows the highest test Dice score as compared to all other baselines and random selection of data for both spleen and prostate. With 24\% of data with the proxy data selection method can achieve up to 90\% of the performance of the full dataset. 

\begin{figure}[!htb]
\begin{center}
    \includegraphics[width=\linewidth]{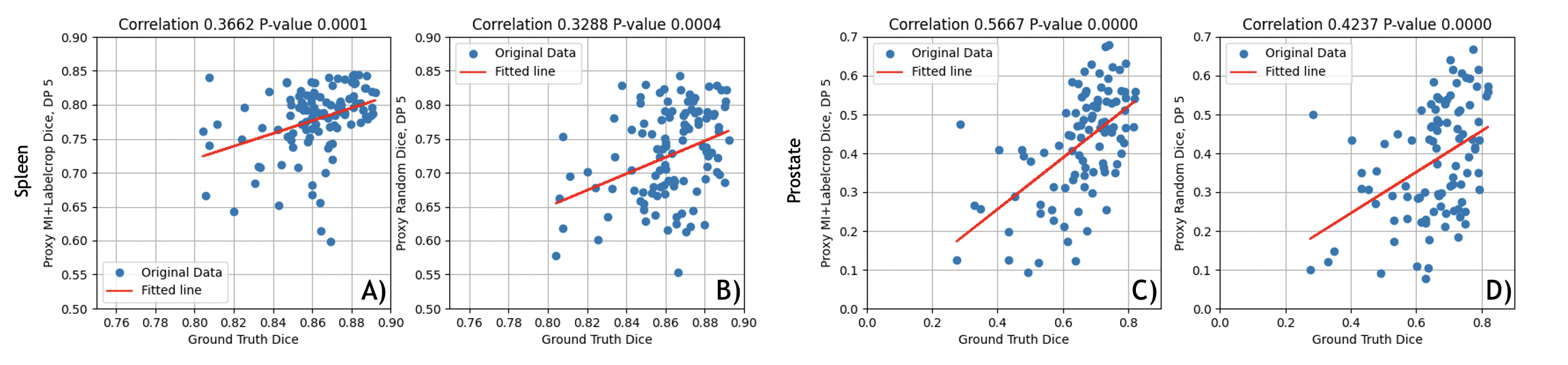}
    \caption{Across 7 splits with different seeds. A) The ``upper bound" Dice of BTCV external dataset is reported when trained on all MSD data. This is compared with Dice on BTCV when trained on a proxy dataset across the hyper-parameter space. B) Similarly, random data is used instead of proxy data. 24 \% of full data were used respectively for A) \& B). C) The ``upper bound" Dice on PROSTATEx is reported when trained on all MSD data. This is compared with Dice on PROSTATEx when trained on a proxy dataset across the hyper-parameter space. D) Similarly, random data is used instead of proxy data. 31\% of all data were used respectively for C) \& D).}
    \label{fig2:proxy_data_spleen}
\end{center}
\end{figure}

Observing across the entire hyper-parameter space the proxy data selection method \textit{MI+Labelcrop} shows a higher correlation of 0.37 versus 0.32 as compared to random selection of data on the external validation when being compared with the ground truth which is obtained by training with all the data (shown in Fig.~\ref{fig2:proxy_data_spleen}).

Similarly, for prostate segmentation, the best test Dice score against the ground truth across the seven repeats of the hyper-parameter optimization was used to detect the best hyper-parameter setting. The test Dice scores reported in (listed in Tab.~\ref{tab:method_select}) belong to the same hyper-parameter setting. The best performing data selection method is \textit{MI+Labelcrop} as also for Spleen. The correlation for \textit{MI+Labelcrop} is in the high range as compared to random which is in the moderate range of correlation.

\textbf{How much data is proxy enough?}
It can be observed that for spleen the cross-over for proxy enough begins to start showing at approximately 15\% of data being used (Fig. \ref{fig5:how_much_data_automl_hp}A \& \ref{fig5:how_much_data_automl_hp}B). A similar observation can be made for prostate as well.

\begin{figure}[!htb]
\begin{center}
    \includegraphics[width=0.7\linewidth]{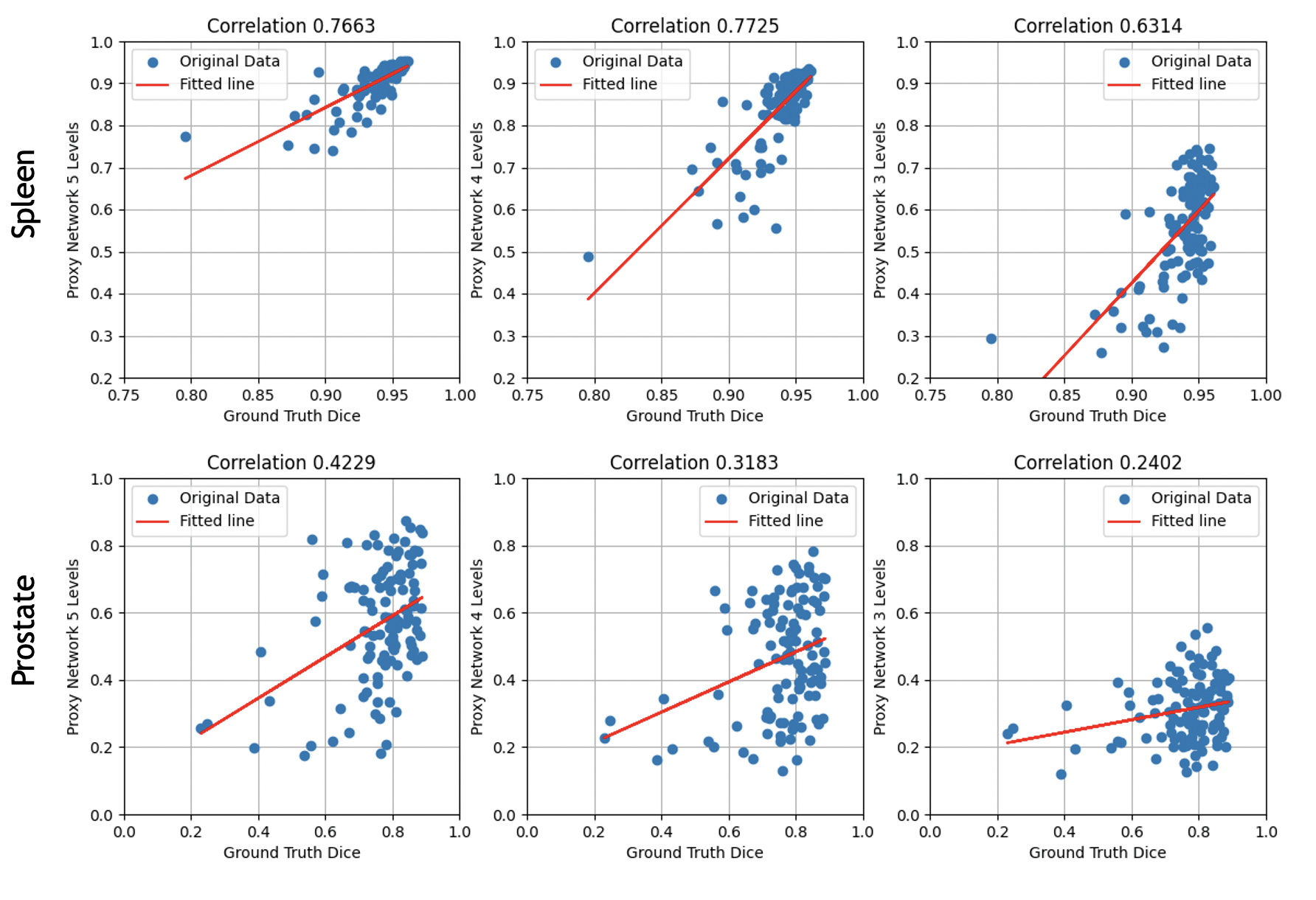}
    \caption{Top row(Spleen): Proxy network correlation when the Dice score on the internal validation of the ground truth versus Dice score from the proxy network across the full data. From left to right reduce the number of levels of the U-net from 5 to 3 (Left to right corresponding levels are 5,4 and 3). All proxy U-nets have 4 channels and one residual block per level. Bottom row (Prostate): Number of levels of U-net are varied and the correlation is compared on the internal validation for prostate data similarly as for spleen.}
    \label{fig4:proxy_network}
\end{center}
\end{figure}

\begin{figure}[!htb]
\begin{center}
    \includegraphics[width=\linewidth]{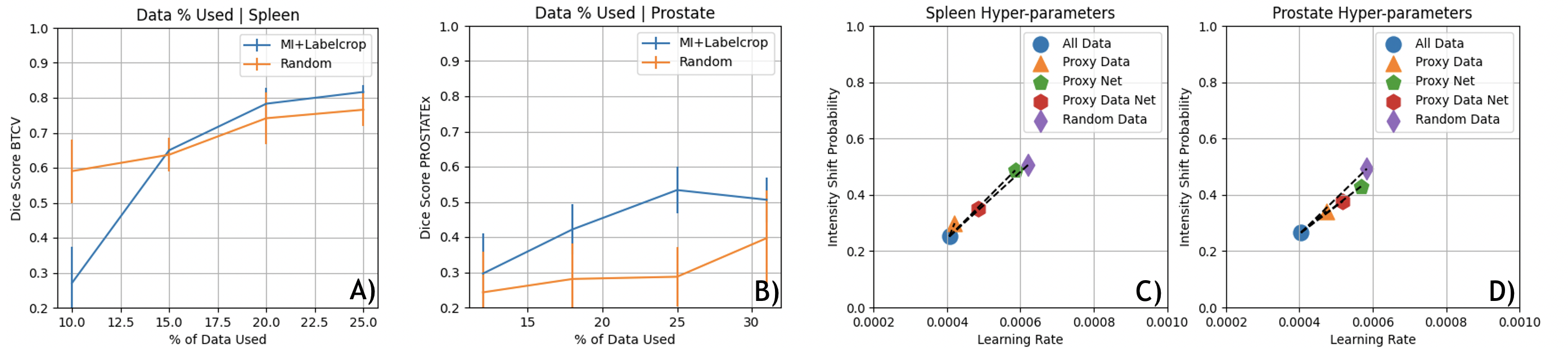}
    \caption{A): For spleen BTCV dataset Dice score is plotted versus data usage. B): For PROSTATEx dataset Dice score is plotted versus data usage. C) For spleen the estimated hyper-parameter of learning rate and probability of intensity shift are shown with relative distance to the ground truth. D) Similarly for prostate}
    \label{fig5:how_much_data_automl_hp}
\end{center}
\end{figure}

\textbf{Proxy Networks:} Across the 7 different splits of data when the validation Dice score is compared for a proxy network versus the full network, a decreasing trend of correlation can be seen (shown in Fig.~\ref{fig4:proxy_network} top row). Given that the channels were fixed at 4 and only a single residual block was used, the correlation is highest for 5 levels in the U-net. Decreasing the number of levels of U-Net decreases the correlation. For all 3 variants a high degree ($>0.5$ ) of correlation is shown, suggesting that even smaller proxy networks could be utilized.

A similar observation as for spleen can be assessed for prostate across the 7 different splits of data. The validation Dice score is compared for a proxy network versus the full network, a decreasing trend of correlation can be seen (shown in Fig.~\ref{fig4:proxy_network} bottom row).

\textbf{AutoML:} The estimated hyper-parameters by RL using proxy data and proxy networks are closer to the full model with all data relatively as compared to when using a random subset of data with the full model for spleen (Fig.~\ref{fig5:how_much_data_automl_hp}C \& \ref{fig5:how_much_data_automl_hp}D). A similar observation can be made for prostate.

For prostate, all proposed proxy methods perform similarly or higher (proxy data and network) vs. random selection of data or when all data is utilized for hyper-parameter estimation (Tab.~\ref{tab:automl}). A similar observation can be made for spleen where proxy data provides the best results. The other methods perform similarly. Overall, a runtime improvement of 4.4$\times$ for spleen and 3.3$\times$ for prostate can be observed.

\begin{table*}[!ht]
  \centering
  \scriptsize
  \begin{tabular}{p{1.8cm} p{2.0cm} p{2.0cm} p{2.0cm} p{2.0cm} p{2.0cm} }
     \hline
     %\multicolumn{6}{c}{\textbf{Proxy Data + Proxy Networks + AutoML}} \\
     \hline
     HP from: & All Data & Proxy Data & Proxy Net & Proxy Data\&Net & Random Data   \\
     \hline
      \multicolumn{6}{c}{\textbf{Prostate}} \\
     MSD Dice & $0.8763 \pm 0.02143$ & $0.8798 \pm 0.0182$ & $0.8764 \pm 0.01883$ & \textbf{0.8812} $\pm 0.0087$ & $0.8204 \pm 0.0346$ \\
     PRSTx Dice & $0.7465 \pm 0.0309$ & $0.7462 \pm 0.0511$ & $0.7493 \pm 0.0334$ & \textbf{0.7701} $\pm 0.0249$ & $0.7134 \pm 0.0353$   \\
     GPU Hours & 772 & 480 & 432 & \textbf{248} & 480  \\
     \hline
     \multicolumn{6}{c}{\textbf{Spleen}} \\
     MSD Dice & $0.9532 \pm 0.0010$ & \textbf{0.9546} $\pm 0.0011$ & $0.9537 \pm 0.0010$ & $0.9536 \pm 0.0011$ & $0.9534 \pm 0.0011$  \\
     BTCV Dice & $0.8780 \pm 0.0103$ & \textbf{0.8821} $\pm 0.0085$ & $0.8763 \pm 0.0087$ & $0.8798 \pm 0.0048$ & $0.8780 \pm 0.0104$  \\
     GPU Hours & 1056 & 320 & 282 & \textbf{240} & 320  \\
     \hline
     
     \hline
  \end{tabular}
  \caption{Summarized mean Dice score for internal validation of MSD and external validation of PROSTATEx and BTCV across 5 repeats with the same split. The estimated hyperparameters (HP) are used to train full models with all data from MSD for comparison.}
  \label{tab:automl}
\end{table*}

\section{Discussion \& Conclusions}
We show that proxy data and proxy networks are a powerful tool to speed up the HPO estimation process. The results indicate that a maximum speedup of 4.4$\times$ can be obtained which can reduce days to a few hours. As a limitation the pairwise distance measures lead to a squared run-time, however the squared runtime is feasible for datasets in the size of thousands, which so far is uncommon in medical imaging segmentation tasks. While this work is a first step towards utilization of proxy techniques for basic hyper-parameter estimation in medical image segmentation, in future we plan to extend it for estimation of multiple hyper-parameters. The benefits can be extended towards multiple frameworks such as neural architecture search and federated learning which are both resource-critical settings. 

\bibliographystyle{splncs04}
\bibliography{egbib}

%Author, F.: Article title. Journal \textbf{2}(5), 99--110 (2016)

%Author, F., Author, S.: Title of a proceedings paper. In: %Editor,
%F., Editor, S. (eds.) CONFERENCE 2016, LNCS, vol. 9999, pp. %1--13.
%Springer, Heidelberg (2016). \doi{10.10007/1234567890}

%Author, F., Author, S., Author, T.: Book title. 2nd edn. %Publisher,

%Author, A.-B.: Contribution title. In: 9th International Proceedings
%on Proceedings, pp. 1--2. Publisher, Location (2010)

%LNCS Homepage, \url{http://www.springer.com/lncs}. Last accessed 4
%\end{thebibliography}
\end{document}